\documentclass[12pt]{iopart}

\usepackage[english]{babel}
\usepackage[ansinew]{inputenc}
\usepackage{iopams}

\newcommand{\dd}[2]{\frac{\mathrm{d} #1}{\mathrm{d} #2}}
\newcommand{\pdl}[2]{\frac{\partial #1}{\partial #2}}
\newcommand{\pds}[2]{\partial #1 / \partial #2}
\newcommand{\ddl}[2]{\frac{\delta #1}{\delta #2}}
\newcommand{\ddd}{\mathrm{d}}

\newcommand{\nn}{\mathbf{\nabla}}

\begin{document}
\title[Rayleigh-Bénard Convection as a Nambu-metriplectic problem]{Rayleigh-Bénard Convection as a Nambu-metriplectic problem}
\author{A Bihlo}
\address{University of Vienna, Institute of Meteorology and Geophysics, Althanstr. 14, 1090 Vienna, Austria}
\ead{alexander.bihlo@univie.ac.at}
\begin{abstract}
The traditional Hamiltonian structure of the equations governing conservative Rayleigh-Bénard convection (RBC) is singular, i.e. it's Poisson bracket possesses nontrivial Casimir functionals. We show that a special form of one of these Casimirs can be used to extend the bilinear Poisson bracket to a trilinear generalised Nambu bracket. It is further shown that the equations governing dissipative RBC can be written as the superposition of the conservative Nambu bracket with a dissipative symmetric bracket. This leads to a Nambu-metriplectic system, which completes the geometrical picture of RBC.
\end{abstract}
\submitto{J. Phys. A: Math. Theor.}

\section{Introduction}

The noncanonical Hamiltonian form of the hydro-thermodynamical equations in Eulerian variables is typically singular. This gives rise to the existence of a special class of conserved quantity, the Casimir functionals. This singularity is a consequence of the reduction that takes place if one changes the coordinates from the (canonical) Lagrangian coordinates to the (noncanonical) Eulerian coordinates by means of the particle-relabeling symmetry (e.g. Thiffeault and Morrison 2000).

In the noncanonical Hamiltonian form the existence of additional conserved quantities is still hidden and hence it is natural to seek for a formulation which lets enter them the description in a similar way as the Hamiltonian does. For the great majority of hydrodynamical systems this is possible (Névir and Blender 1993, Névir 1998) and leads to a form of description that formally resembles the structure of Nambu mechanics, which was first introduced by Nambu (1973) for discrete systems. 

We start with the classical definition of a discrete Nambu system (Takhtajan 1994):

\medskip

\noindent{\bf Definition:} Let $M$ be a smooth manifold and $C^{\infty}(M)$ the algebra of infinitely differentiable real valued functions defined on $M$. Then $M$ is called a \textit{Nambu-Poisson manifold} of order $n$ if there exists a map $\{,\dots,\}: [C^\infty(M)]^{\otimes n} \to C^\infty(M)$ called the \textit{Nambu bracket} that satisfies
\begin{enumerate}
	\item $\{f_1,f_2,\dots,f_n\} = (-1)^{\epsilon(\sigma)}\{f_{\sigma(1)},f_{\sigma(2)},\dots,f_{\sigma(n)}\}$
	where $\sigma$ denotes an element of the symmetric group of $n$ elements, with $\epsilon(\sigma)$ being it's parity.
	\item $\{f_1f_2,f_3,\dots,f_{n+1}\} = f_1\{f_2,f_3,\dots,f_{n+1}\}+\{f_1,f_3,\dots,f_{n+1}\}f_2$
	\item $\{\{f_1,\dots,f_{n-1},f_{n}\},f_{n+1},\dots,f_{2n-1}\}+\{f_n,\{f_1\dots,f_{n-1},f_{n+1}\},f_{n+2},\dots,f_{2n-1}\}$ $+$ $\cdots +\{f_n,\dots,f_{2n-2},\{f_{1},\dots, f_{n-1},f_{2n-1}\}\} = \{f_1,\dots,f_{n-1},\{f_{n},\dots,f_{2n-1}\}\}$
\end{enumerate}
Note that the first property defines the skew-symmetry of the Nambu bracket, the second property is the Leibniz rule and the last one is a generalisation of the Jacobi identity, known as the Fundamental identity (FI) or Takhtajan identity.

The Nambu bracket defines the kinematic part of a Nambu system, which is supplemented by $n-1$ function $H_1,\dots, H_{n-1}$, such that the evolution of a real-valued function on $M$ is given by

\[
	\dd{f}{t} = \{f,H_1,\dots,H_{n-1}\}.
\]
Due to the antisymmetry of the Nambu bracket, it follows that $H_1,\dots,H_{n-1}$ are conserved by the Nambu flow.

The generalisation to field equations was done by Névir and Blender (1993) and Névir (1998) by starting with the noncanonical Hamiltonian formulation of the respective model equations. They extended them by means of using \emph{one} of their Casimirs as additional conserved quantity. Hence, their continuous Nambu formulation also only uses \emph{trilinear} bracket structures. This kind of generalisation is the one we aim to use for the equations governing RBC.

\medskip

\noindent Let us now turn to dissipative systems with a conservative Hamiltonian core. Although the notion of such so called metriplectic systems is not unique (see Guha 2007 for a review) the constituent parts of them are the antisymmetric Poisson and a symmetric (or gradient) structure, that accounts for dissipation. Adding both pieces together then describes the dynamics of the whole dissipative system. The metriplectic bracket hence reads

\[
	\langle \! \langle f,g \rangle \! \rangle \,:= \{f,g\} + \langle f, g \rangle
\]
where
\begin{eqnarray}
	\{f,g\} = -\{g,f\} \qquad\qquad \textnormal{antisymmetric (Poisson) bracket} \nonumber \\
	\langle f,g \rangle = \langle g,f \rangle \qquad\qquad\quad\!\,\, \textnormal{symmetric (gradient) bracket} \nonumber 
\end{eqnarray}
holds.

In this paper we show that the equations of RBC can be cast in Nambu-metriplectic form, that is, the conservative part possesses a Nambu and the dissipative part a symmetric bracket structure.

\section{The Nambu structure of conservative RBC}

The equations of two-dimensional RBC in case of an incompressible fluid using the Boussinesq approximation in nondimensional form read (e.g. Thiffeault and Horton 1996):

\begin{equation} \label{RBC}
	\pdl{\zeta}{t} + [\psi,\zeta] = \pdl{T}{x} + \nu\nn^4\psi, \qquad\qquad
    \pdl{T}{t} + [\psi,T] = \pdl{\psi}{x} + \kappa\nn^2 T.
\end{equation}
As usual, $\psi$ is the stream function generating two-dimensional nondivergent flow in the $x$--$z$-plane, $\zeta = \nn^2\psi$ is the vorticity, $T$ is the temperature departure of a linear conduction profile, $\nu$ and $\kappa$ are kinematic viscosity and thermal conductivity, respectively. $[a,b] := \pds{a}{x}\,\pds{b}{z}-\pds{a}{z}\,\pds{b}{x}$ denotes the Jacobian operator.

In the following, we will assume that the domain of interest is periodic in the $x$-direction. In the vertical, we either also assume periodicity or stress-free boundary conditions (Saltzman, 1962), i.e.
\[
	\psi = 0,\quad \zeta = 0,\quad \pdl{\psi}{x}=0 \qquad \textnormal{at}\qquad z=0,\quad z=z_{\mathrm{top}}.
\]
In either case, since in RBC a constant temperature difference is maintained externally between the top and bottom of the fluid, the appropriate boundary condition for the temperature deviation $T$ is
\[
	T = 0 \qquad \textnormal{at}\qquad z=0,\quad z=z_{\mathrm{top}}.
\]
Determining the classical continuous Nambu form is possible only in the case of vanishing dissipation, i.e. in case $\nu = \kappa = 0$. Then, since the remaining terms on the right hand side can be written as $\pds{T}{x} = [T,z]$ and $\pds{\psi}{x} = [\psi,z]$, respectively, both equations may be arranged as

\begin{equation} \label{startsys}
	\pdl{\zeta}{t} + [\psi,\zeta] + [z, T-z] = 0, \qquad\qquad
    \pdl{T}{t} + [\psi,T-z] = 0.
\end{equation}
This set of equations is Hamiltonian upon using

\[
 \mathcal{H} = \int_\Omega \left(\frac{1}{2}(\nn\psi)^2 - Tz\right)\ddd f
\]
as Hamiltonian functional and

\begin{equation} \label{poisson}
  \fl  \{\mathcal{F},\mathcal{H}\} = \int_\Omega \left(\zeta\left[\ddl{\mathcal{F}}{\zeta},\ddl{\mathcal{H}}{\zeta}\right]+ (T-z)\left( \left[\ddl{\mathcal{F}}{\zeta},\ddl{\mathcal{H}}{T}\right] + \left[\ddl{\mathcal{F}}{T},\ddl{\mathcal{H}}{\zeta}\right] \right)\right) \ddd f
\end{equation}
as a Poisson bracket. An analogue bracket arises also in magnetohydrodynamics (MHD) (Morrison and Hazeltine 1984). Note, however, that this Poisson bracket is singular, i.e. it possesses nonvanishing Casimir functionals. They are

\[
	\mathcal{C}_1 = \int_\Omega g(T-z)\,\ddd f,\qquad\qquad \mathcal{C}_2 = \int_\Omega \zeta h(T-z)\,\ddd f,
\]
where $g,h$ are arbitrary functions of $T-z$. For a physical intepretation of the analogue Casimirs in MHD, see Thiffeault and Morrison (1998). In conservative RBC the first class of Casimirs $\mathcal{C}_1$ physically describes the preservation of $T-z$-contours. In fact, this class of Casimirs generally arises in nondivergent and inviscid fluid models. In turn, the second class $\mathcal{C}_2$ incorporates the Kelvin's circulation theorem, which in the case of RBC requires conservation of the vorticity $\zeta$ on closed contours of $T-z$.

To determine the conservative Nambu form, we are only interested in the special form of $\mathcal{C}_2$ with $h=T-z$:
\[
    \mathcal{C} = \int_\Omega\zeta(T-z)\,\ddd f. \nonumber 
\]
It allows us to extend the Poisson bracket formulation of the governing equations to arrive at the their Nambu form:
\begin{eqnarray}
    \pdl{\zeta}{t} = -\left[\ddl{\mathcal{C}}{T},\ddl{\mathcal{H}}{\zeta}\right] - \left[\ddl{\mathcal{C}}{\zeta},\ddl{\mathcal{H}}{T}\right] &=& \{\zeta,\mathcal{C},\mathcal{H}\} \nonumber \\
    \pdl{T}{t} = -\left[\ddl{\mathcal{C}}{\zeta},\ddl{\mathcal{H}}{\zeta}\right] &=& \{T,\mathcal{C},\mathcal{H}\}, \nonumber
\end{eqnarray}
where the bracket $\{\cdot,\cdot,\cdot\}$ is defined for arbitrary functionals $\mathcal{F}_1$, $\mathcal{F}_2$ and $\mathcal{F}_3$ by the equation
\begin{eqnarray} \label{nambu}
	\fl \{\mathcal{F}_1,\mathcal{F}_2,\mathcal{F}_3\} := -\int_\Omega \Bigg( \ddl{\mathcal{F}_1}{T}\left[\ddl{\mathcal{F}_2}{\zeta},\ddl{\mathcal{F}_3}{\zeta}\right]+ \ddl{\mathcal{F}_1}{\zeta}\left[\ddl{\mathcal{F}_2}{T},\ddl{\mathcal{F}_3}{\zeta}\right] +\ddl{\mathcal{F}_1}{\zeta}\left[\ddl{\mathcal{F}_2}{\zeta},\ddl{\mathcal{F}_3}{T}\right]\Bigg) \ddd f.
\end{eqnarray}
This bracket is easily seen to be totally antisymmetric in case we assume periodic boundary conditions in both directions. Another possibility that guarantees antisymmetry of (\ref{nambu}) is provided by periodicity in $x$-direction and free boundaries in the vertical. For this choice, however, it is necessary to fix $\mathcal{F}_3=\mathcal{H}$ in the Nambu bracket. This is explicitly shown in the appendix. Assuring antisymmetry is also possible by only considering those class of functionals that sufficiently rapidly go to zero towards the boundaries. All these choices guarantee that surface terms emerging from an integration by parts vanish.

The FI is proved by noting that the above Nambu bracket is simply the continuous analogue of the heavy top Nambu bracket. Hence, this bracket may indeed serve as a good Nambu bracket.

Note that $\mathcal{C}$ is indefinite with respect to sign. In this respect it is akin to the helicity, which is a Casimir for the three dimensional incompressible Euler equations (Névir and Blender, 1993). To our knowledge, the latter model and RBC are the only known ones that need indefinite Casimirs to allow for a Nambu representation.

\section{The Nambu-metriplectic structure of dissipative RBC}

Now turning to the dissipative equations (i.e. $\nu\ne 0, \kappa\ne 0$), we aim to show that this model has Nambu-metriplectic form. Let us first note, that adding $0=[T-z,T-z]$ to the first equation in (\ref{startsys}) leads to the equivalent system

\[
	\pdl{\zeta}{t} + [\psi,\zeta] + [T, T-z] = 0, \qquad\qquad
    \pdl{T}{t} + [\psi,T-z] = 0.
\]
which is Hamiltonian upon using (\ref{poisson}) as Poisson bracket and

\[
	\mathcal{G} = \int_\Omega \left(\frac{1}{2}(\nn\psi)^2 - Tz - \frac{1}{2}(T-z)^2\right)\ddd f = \mathcal{H} - \mathcal{S} 
\]
as Hamiltonian functional. Note that $\mathcal{S}$ is the particular realisation of the first class of Casimir functionals $\mathcal{C}_1$ with $g = 1/2(T-z)^2$. In Morrison (1986) functionals like $\mathcal{G}$ are termed generalised free energy.

This formulation enables us to cast the equations governing dissipative RBC in Nambu-metriplectic form. Indeed, in this case the equations can be written as
\begin{eqnarray}
	\pdl{\zeta}{t} + \left[\ddl{\mathcal{C}}{T},\ddl{\mathcal{G}}{\zeta}\right] + \left[\ddl{\mathcal{C}}{\zeta},\ddl{\mathcal{G}}{T}\right] = -\nu\nn^4\ddl{\mathcal{G}}{\zeta} \nonumber \\
	\pdl{T}{t} + \left[\ddl{\mathcal{C}}{\zeta},\ddl{\mathcal{G}}{\zeta}\right] = -\kappa\nn^2\ddl{\mathcal{G}}{T} \nonumber
\end{eqnarray}
and it is possible to introduce an indefinite symmetric bracket that governs dissipation:

\begin{equation}\label{diss}
	\left\langle \mathcal{F}, \mathcal{G}\right\rangle := -\int_\Omega\left(\nu\ddl{\mathcal{F}}{\zeta}\nn^4\ddl{\mathcal{G}}{\zeta} + \kappa\ddl{\mathcal{F}}{T}\nn^2\ddl{\mathcal{G}}{T}\right)\ddd f.
\end{equation}
The symmetry property of this bracket is assured if either periodic boundaries are assumed or only functionals that sufficiently rapidly decay to zero near the boundaries are considered. Adding together both brackets gives the entire dynamics of two-dimensional RBC:

\[
\pdl{\zeta}{t} = \{\zeta,\mathcal{C},\mathcal{G}\} + \left\langle\zeta, \mathcal{G}\right\rangle,
    \qquad\qquad
    \pdl{T}{t} = \{T,\mathcal{C},\mathcal{G}\} + \left\langle T, \mathcal{G}\right\rangle.	
\]
Note that in some sense the Nambu-metripectic formulation of (\ref{RBC}) is geometrically the most complete, since representatives of \emph{both} classes of Casimirs are needed to represent the whole dynamics.

The existence of dissipation generally spoils the conservation properties of conservative system. This is also the case in RBC, since we have

\[
	\pdl{\mathcal{G}}{t} = \left\langle \mathcal{G}, \mathcal{G}\right\rangle \ne 0, \qquad\qquad \pdl{\mathcal{C}}{t} = \left\langle \mathcal{C}, \mathcal{G}\right\rangle \ne 0.
\]
That is, the evolution of $\mathcal{G}$ is determined solely by $\mathcal{G}$, whereas the evolution of $\mathcal{C}$ is determined both by $\mathcal{C}$ and $\mathcal{G}$.

\section{Comments and outlook}

In this work we have shown that the conservative part of (\ref{RBC}) can be written in Nambu form, while the full dissipative system possesses a Nambu-metriplectic form. For both representations, the explicit usage of Casimir functionals of (\ref{poisson}) is crucial. The Casimir $\mathcal{C}$ allows to extend the bilinear Poisson bracket to a trilinear Nambu bracket. In turn, the Casimir $\mathcal{S}$ can be subtracted from the Hamiltonian $\mathcal{H}$ to give the modified Hamiltonian $\mathcal{G}$. This doesn't alter the dynamics, since Casimirs are trivial conserved quantities and Hamiltonians are only determined up to Casmir functionals. But introducing $\mathcal{G}$ is essential to allow for the necessary symmetry property of the dissipative bracket (\ref{diss}).

The Poisson bracket (\ref{poisson}) is an example of a Lie-Poisson system. Such systems are built from an underlying Lie algebra structure, typically owing to a reduction from a set of canonical to a set of noncanonical variables (e.g. Thiffeault and Morrison 2000). In case of RBC it is the semi-direct extension of the algebra associated to the group of volume-preserving diffeomorphisms on some domain $\Omega$ with the vector space of real-valued functions on $\Omega$ (Thiffeault and Morrison 2000). As was shown in our work, the Nambu structure is compatible with this type of algebra extension. In particular, the Nambu bracket (\ref{nambu}) allows to put (\ref{poisson}) in a more symmetric form.

Studying algebras and extensions thereof in conjunction with Nambu structures offers a way to classify Nambu systems. Such work is currently in progress and will be published elsewhere.

\ack The author thanks Dr. Roman Popovych for the numerous essential discussions. The useful hints by the anonymous referee, which lead to a much improved version of this work are gratefully acknowledged.

\appendix
\section*{Appendix}

We aim to explicitly show here that assuming periodic boundaries in $x$-direction and free boundaries in the vertical guarantees the total antisymmetry of (\ref{nambu}) provided we fix the Hamiltonian $\mathcal{H}$ in the bracket.

We only need to prove antisymmetry with respect to the first two functionals of the bracket (\ref{nambu}). The antisymmetry in the other pair of arguments is obvious. An integration by parts gives
\begin{eqnarray}
	\int_\Omega \ddl{\mathcal{F}}{T}\left[\ddl{\mathcal{C}}{\zeta},\ddl{\mathcal{H}}{\zeta}\right]\ddd f = -\int_\Omega\ddl{\mathcal{C}}{\zeta} \left[\ddl{\mathcal{F}}{T},\ddl{\mathcal{H}}{\zeta}\right]\ddd f - \int_L\left(\ddl{\mathcal{F}}{T}\ddl{\mathcal{C}}{\zeta}\pdl{}{x}\ddl{\mathcal{H}}{\zeta}\right) \Big|_z\ddd x\nonumber \\
	\int_\Omega \ddl{\mathcal{F}}{\zeta}\left[\ddl{\mathcal{C}}{T},\ddl{\mathcal{H}}{\zeta}\right]\ddd f = -\int_\Omega\ddl{\mathcal{C}}{T} \left[\ddl{\mathcal{F}}{\zeta},\ddl{\mathcal{H}}{\zeta}\right]\ddd f - \int_L\left(\ddl{\mathcal{F}}{\zeta}\ddl{\mathcal{C}}{T}\pdl{}{x}\ddl{\mathcal{H}}{\zeta}\right) \Big|_z\ddd x\nonumber \\
		\int_\Omega \ddl{\mathcal{F}}{\zeta}\left[\ddl{\mathcal{C}}{\zeta},\ddl{\mathcal{H}}{T}\right]\ddd f = -\int_\Omega\ddl{\mathcal{C}}{\zeta} \left[\ddl{\mathcal{F}}{\zeta},\ddl{\mathcal{H}}{T}\right]\ddd f - \int_L\left(\ddl{\mathcal{F}}{\zeta}\ddl{\mathcal{C}}{\zeta}\pdl{}{x}\ddl{\mathcal{H}}{T}\right) \Big|_z\ddd x\nonumber
\end{eqnarray}
were we have already taken into account periodicity in the $x$-direction. Due to the imposed boundary conditions

\[
	\pdl{}{x}\left(\ddl{\mathcal{H}}{\zeta}\right)\Big|_z = -\pdl{\psi}{x}\Big|_z = 0, \qquad \pdl{}{x}\left(\ddl{\mathcal{H}}{T}\right)\Big|_z = -\pdl{z}{x}\Big|_z = 0
\]
all the second terms on the right hand side vanish. Similar considerations also hold for fixing the modified Hamiltonian $\mathcal{G}$.

\section*{References}

\begin{harvard}

\item[] Guha P 2007 {\it J. Math. Anal. Appl.} {\bf 326}, 121
\item[] Morrison P J and Hazeltine R D 1984 {\it Phys. Fluids} {\bf 27}, 886
\item[] Morrison P J 1986 {\it Physica D} {\bf 18} 410
\item[] Nambu Y 1973 {\it Phys. Rev. D} {\bf 7} 2405
\item[] Névir P and Blender R 1993 {\it J. Phys. A} {\bf 26} L1189
\item[] Névir P 1998 Die Nambu-Felddarstellung der Hydro-Thermodynamik und ihre Bedeutung f\"ur die dynamische Meteorologie (Freie Universit\"at Berlin) {\it Habilitation thesis}
\item[] Saltzman B 1962 {\it J. Atmos. Sci.} {\bf 19} 329
\item[] Takhtajan L 1994 {\it Comm. Math. Phys.} {\bf 160} 295
\item[] Thiffeault J L and Morrison P J 1998 {\it Ann N Y Acad Sci.} {\bf 867} 109 arXiv:chao-dyn/9804032v1
\item[] Thiffeault J L and Morrison P J 2000 {\it Physica D} {\bf{136}}, 205
\item[] Thiffeault J L and Horton W 1996 {\it Phys. Fluids} {\bf 8} 1715 arXiv:chao-dyn/9603019v1

\end{harvard}

\end{document}